\documentclass[aps,pra,amsmath,amssymb, twocolumn,superscriptaddress]{revtex4-1}

\usepackage[T1]{fontenc}
\usepackage{amssymb,amsmath}
\usepackage{amsfonts}
\usepackage{graphicx,color,float}
\usepackage{amsthm,hyperref}
\usepackage{lineno}
 \usepackage{soul}   				

\def\ket#1{|#1\rangle}
\def\bra#1{\langle#1|}

\def\ketbra#1{\left|#1\rangle\langle#1\right|}

\def\tr{\mathrm{tr}}

\definecolor{mygray}{gray}{0.6}

\newcommand{\SD}{S_{\rm D}}
\newcommand{\Sde}{S_{_{\overline{\rho(\tau)}}}}

\newcommand{\ktimes}{\rangle\! \langle}
\newcommand{\op}[2]{|#1\ktimes #2|}
\newcommand{\equa}[1]{Eq.~(\ref{#1}) }

\newcommand{\ifimar}{Instituto de Investigaciones F\'isicas de Mar del Plata  (CONICET-UNMdP),  B7602AYL Mar del Plata,  Argentina}
\newcommand{\conicet}{Consejo Nacional de Investigaciones Cient\'ificas y Tecnol\'ogicas (CONICET), C1425FQB C.A.B.A, Argentina}
\begin{document}
\title{Average diagonal entropy in non-equilibrium isolated quantum systems}
\author{Olivier Giraud}
\affiliation{LPTMS, CNRS, Univ.~Paris-Sud, Universit\'e Paris-Saclay, 91405 Orsay, France}
\author{Ignacio Garc\'ia-Mata} \affiliation{\ifimar} \affiliation{\conicet}
\begin{abstract}
The diagonal entropy was introduced as a good entropy candidate especially for isolated quantum systems out of equilibrium. 
Here we present an analytical calculation of the average diagonal entropy for systems undergoing unitary evolution and an external perturbation in the form of a cyclic quench. We compare our analytical findings with numerical simulations of various many-body quantum systems. 
Our calculations elucidate various heuristic relations proposed recently in the literature.
\end{abstract}
\maketitle

The precision for manipulating quantum systems attained to date 
has led to the next big question: how do basic thermodynamics principles operate at very small scales. 
Motivated by this question
an ever growing effort has surged attempting to describe quantum thermodynamics of small isolated quantum systems, and their approach to equilibrium and subsequent thermalization. The incredible advance in experimental techniques allowing to follow the time evolution of closed quantum systems
 \cite{Paredes2004,*Kinoshita,*Hofferbert} has been the main boost of these endeavors. The relevance of this subject is 
 (at least) twofold. On the one hand, quantum technologies (e.g.~for quantum information 
 \cite{BlattRMP} and quantum simulation  \cite{Gerri,*Blatt2012,*Serwane,*Korenblit}) tend to be based on systems with negligible interaction with the environment. On the other hand, a complete microscopic thermodynamical description of these advances in nonequilibrium statistical mechanics of such quantum systems has remained elusive, in part
  due to the lack of a suitable definition of entropy.
Although the von Neumann entropy $S_{\rm vN}=-\tr \rho \ln \rho$ (with $k_{\rm B}=1$)  is a natural tool to measure the entropy of a quantum state $\rho$, it cannot be used to describe the approach to equilibrium of isolated quantum systems, since they undergo unitary dynamics. As an alternative the diagonal entropy (DE)
\begin{equation}
\label{defDE}
\SD=-\sum_n \rho_{nn}\ln\rho_{nn}
\end{equation}
was proposed \cite{Polkov2011}, where $\rho_{nn}$ are the diagonal elements of the density matrix in the energy eigenbasis. 
The DE possesses most of the expected features of a thermodynamic entropy, such as additivity, or increase when a system at equilibrium undergoes an external perturbation~\cite{Polkov2011, Ikeda2015}. The DE appears to be a fundamental quantity, which can describe the behavior of a very broad class of out-of-equilibrium systems. Clarifying the universality or the specificities of its properties is therefore an important goal. 

Two universal properties of DE have been proposed recently. In the case where a system is perturbed by an external operation during a time $\tau$, one can study the DE as a function of the duration $\tau$ of the perturbation. In \cite{Ikeda2015} a conjecture was made introducing bounds on the difference between the time averaged DE, $\overline{\SD(\tau)}$, and the DE of the time averaged state $\overline{\rho(\tau)}$  (denoted as $\Sde$), namely
\begin{equation}
\label{ikeda}
0 \le \Delta S\le 1- \gamma, \ \text{ where }\ \Delta S \stackrel{\rm def}{=}\Sde-\overline{\SD(\tau)}
\end{equation}
and  $\gamma=0.5772\ldots$ is Euler's constant. A second property was numerically uncovered in \cite{GarciaMata2015}, where a universal relation was found between $\Delta S$ and a quantity 
measuring localization of the initial state. 

The aim of this Letter is to provide analytical support to both of these observations and to determine to which extent they are universal. To this end, we derive an analytical expression for $\Delta S$ as an expansion in terms of average generalized participation ratios (PR), which characterize the localization properties of the vector of transition probabilities between the initial and the perturbed eigenstates. Using perturbation theory, we analyze the behavior of the first terms of this expansion in the two extreme regimes of localized and delocalized states.
To support our analytical findings we performed numerical simulations using two representative physical models displaying chaotic and integrable regimes.  
We show that our truncated expansion of $\Delta S$ is accurate independently of the physical model and of the localization 
properties of the eigenfunctions.

The importance of our results is twofold. 
They provide a  precise analytical value for the time average of $\SD$ and $\Delta S$, improving \cite{Ikeda2015}, and giving a tighter bound.
 Moreover, they relate the DE to localization properties of eigenstates of the perturbed system by an explicit and accurate expression. This should lead to a better understanding of the deep connections between Anderson-type  transitions 
 and equilibration characterized by the DE. 
It is noteworthy that the DE is uniquely related to the energy distribution \cite{Polkov2011} and it is thus (in principle) a measurable quantity. Therefore, it could be relevant in the experimental description of equilibration and thermalization processes which have gained so much attention due to a flurry of recent breakthroughs (see \cite{Jensen85,Deutsch91, Srednicki1994,Calabrese2006,Rigol2008,Rigol2009,Linden2009,GogolinMullerEisert2011} to name but a few).

For simplicity we  consider a cyclic external operation, where a system, described by a Hamiltonian $H=H(\lambda)$ depending on a fixed parameter $\lambda$, undergoes a sudden quench $H\rightarrow H'=H(\lambda+\delta\lambda)$ at time $t=0$ and is then reverted to the Hamiltonian $H$ at time $\tau$. If $H$ is  time-independent, the DE $S_D(t)$ of the state $\rho(t)$ is constant for $t>\tau$. It can thus be studied as a function of the perturbation duration $\tau$, and denoted $S_D(\tau)$. For large enough $\tau$, the system evolving under Hamiltonian $H'$ will have the time to equilibrate, so that $S_D(\tau)$ goes to a constant value that can be estimated by considering the average $\overline{\SD(\tau)}$. 
Let us label by $\ket{n}$ the basis of normalized eigenvectors of $H$ (with eigenvalues $E_n$), and by $\ket{m}$ the basis of eigenvectors of $H'$ (with eigenvalues $E'_m$). We may consider finite systems of size $N$, or truncate our matrices to 
 a Hilbert space of dimension $N$. We assume that at $t=0$ the system is in an eigenstate $\ket{n_0}\bra{n_0}$ of $H$ with energy $E_{n_0}$. Let $U=e^{-i H'\tau}$ ( $\hbar\stackrel{\rm def}{=} 1$) be the evolution operator from time 0 to $\tau$. At time $\tau$ the state is $\rho(\tau)=U\ket{n_0}\bra{n_0}U^{\dagger}$. Its DE can  be expressed as 
\begin{equation}
\label{dent}
S_D(\tau)=-\sum_n h_n\ln h_n,\qquad h_n=|\bra{n}U\ket{n_0}|^2,
\end{equation}
where $h_n$ is the probability to observe the system in an eigenstate $\ket{n}$ at time $\tau$ (we have the normalization $\sum_n h_n=1$). 
Our aim is to calculate the quantity
\begin{equation}
\label{DeltaS}
\Delta S=-\sum_n \overline{h_n}\ln\overline{h_n}+\sum_n \overline{h_n\ln h_n}.
\end{equation}
Since $h_n\in [0,1]$ has finite support, the knowledge of all its moments uniquely defines its distribution $P(h_n)$, from which it is possible to calculate $\Delta S$. Using
\begin{equation}
\label{Unn0}
\bra{n}U\ket{n_0}=\sum_me^{-iE'_m \tau}\langle n\ketbra{m}n_0\rangle,
\end{equation}
the probabilities $h_n$ can be expressed as
\begin{equation}
\label{hnbis}
h_n=\!\!\sum_{m_1,m_2}\!e^{i(E'_{m_1}-E'_{m_2}) \tau}\langle n\ketbra{m_2}n_0\rangle\langle n_0\ketbra{m_1}n\rangle.
\end{equation}
Moments of $h_n$ are obtained by calculating the averages $\overline{h_n^k}$. From Eq.~\eqref{hnbis}, these averages involve averages of quantities $\exp \left[i\left(\sum_{i=1}^k E'_{m_{2i-1}}-\sum_{i=1}^{k} E'_{m_{2i}}\right)\tau\right]$. We make, as in \cite{Ikeda2015}, the assumption that these quantities average to 1 if the sets $\{m_{2i-1},1\leq i\leq k\}$ and $\{m_{2i},1\leq i\leq k\}$ are permutations of each other, and to 0 otherwise. From Eq.~\eqref{hnbis}, keeping only terms with $m_1=m_2$ we have the first moment 
\begin{equation}
\label{hnbar}
\overline{h_n}=\sum_m c_{mn},\qquad c_{mn}=|\langle m|n_0\rangle|^2|\langle m|n\rangle|^2.
\end{equation}
The $\overline{h_n}$ are the average probabilities of transition from $\ket{n_0}$ to $\ket{n}$. For the second moment $\overline{h_n^2}$, keeping only terms for which the sets $\{m_1,m_3\}$ and $\{m_2,m_4\}$ are the same, we get
\begin{equation}
\label{hn2}
\overline{h_n^2}=2\left(\sum_m c_{mn}\right)^2-\sum_m c_{mn}^2.
\end{equation}
For integer $q$, we introduce the PR
\begin{equation}
\label{xiqn}
\xi_{q,n}\equiv\frac{\sum_m c_{mn}^q}{(\sum_m c_{mn})^q},
\end{equation}
which characterize the localization properties of the vectors $(c_{mn})_{m}$. We can then express Eq.~\eqref{hn2} as $\overline{h_n^2}/\overline{h_n}^2=2-\xi_{2,n}$. Higher-order moments can be expressed in the same way. For instance we have $\overline{h_n^3}/\overline{h_n}^3=6-9\xi_{2,n}+4\xi_{3,n}$ and $\overline{h_n^4}/\overline{h_n}^4=24-72\xi_{2,n}+18(\xi_{2,n})^2+64\xi_{3,n}-33\xi_{4,n}$. Keeping only the first two terms in these expressions we get the general formula (see Supplemental material for a detailed proof)
\begin{equation}
\label{hnkordre2}
\overline{h_n^k}=\overline{h_n}^k\, k!\,\left(1-\frac{k(k-1)}{4}\xi_{2,n}\right).
\end{equation}
As $\xi_{2,n}$ does not depend on $k$, the moment generating function of $h_n$ can then be resummed as
\begin{equation}
\label{sumM}
M_n(t)=\sum_{k=0}^{\infty}\frac{\overline{h_n^k}}{k!}t^k=\frac{1}{1-\overline{h_n}t}-\frac{(\overline{h_n}t)^2}{2(1-\overline{h_n}t)^3}\xi_{2,n}.
\end{equation}
The probability distribution for $h_n$ is then obtained by inverse Laplace transform of $M_n(t)$, which gives
\begin{equation}
\label{phnext}
P(h_n)=\frac{1}{\,\overline{h_n}\, }e^{-\tfrac{h_n}{\,\overline{h_n}\,}}\left(1-\left[\frac{h_n^2}{4\overline{h_n}^2}-\frac{h_n}{\,\overline{h_n}\, }+\frac12\right]\xi_{2,n}\right).
\end{equation}
Using this distribution, the calculation of the averages in Eq.~\eqref{DeltaS} is then straightforward and direct integration yields
\begin{equation}
\label{conjecturenext}
\Delta S= 1-\gamma-\frac14\overline{\xi}_{2},\qquad \overline{\xi}_{2}\equiv\sum_n\overline{h_n}\,\xi_{2,n}.
\end{equation}
Recalling that $\sum_n\overline{h_n}=1$, the quantity $\overline{\xi}_{2}$ appears as the average over PRs \eqref{xiqn} weighted by the mean transition probability $\overline{h_n}$ to go from $\ket{n_0}$ to $\ket{n}$ during the quench. This first resul,  Eq.~\eqref{conjecturenext}, has two consequences. First, it makes a more precise statement than the conjecture $\Delta S\leq 1-\gamma$ of \eqref{ikeda} by showing that for finite dimension $N$, as the value of $\overline{\xi}_2$ is finite, the inequality $\Delta S<1-\gamma$ is strictly fullfilled. Moreover, while the equality $\Delta S = 1-\gamma$ was proved in \cite{Ikeda2015} under some assumptions on the localization properties and in the limit $N\to\infty$, our result provides a value for the first correction to the difference between $\Delta S$ and its upper bound in the finite $N$ case.
The second consequence of Eq.~(\ref{conjecturenext}) is that it relates the DE to the average PRs of the overlaps $c_{mn}$, substantiating the observations of \cite{GarciaMata2015}.

In Eq.~\eqref{hnkordre2} we only kept the first two terms in the expression of the moments $\overline{h_n^k}/\overline{h_n}^k$. It is in fact possible to systematically carry out the calculation by keeping successive terms in the expression of the moments, as we show in the Supplemental material. This yields an expansion of the form 
\begin{equation}
\label{bigsum}
\Delta S= 1-\gamma+\sum_{\mu}\frac{a_{\mu}\overline{\xi}_{\mu}}{|\mu|(|\mu|-1)},\quad \overline{\xi}_{\mu}\equiv\sum_n\overline{h_n}\,\xi_{\mu_1,n}\xi_{\mu_2,n}\ldots,
\end{equation}
where $a_{\mu}$ are rational numbers and the sum over $\mu$ runs over all finite integer sequences $\mu=(\mu_1,\mu_2,\ldots)$ such that $\mu_i\geq 2$, and $|\mu|=\sum_i\mu_i$. For instance, keeping the few next lowest-order terms in the expression of moments, \eqref{conjecturenext} can be corrected to 
\begin{eqnarray}
\label{conjecture3}
\Delta S&=& 1-\gamma-\frac14\overline{\xi}_2+\left(\frac19\overline{\xi}_3+\frac{1}{16}\overline{\xi}_{22}\right)\nonumber\\
&&-\left(\frac{11}{96} \overline{\xi}_4+\frac{1}{6} \overline{\xi}_{32}+\frac{1}{16} \overline{\xi}_{222}\right).
\end{eqnarray}
This expression is the main result of this work. It provides a connection between the average DE and the structure of the eigenfunctions -- localization on the perturbed basis -- through the generalized PR \eqref{xiqn}. It is all the more accurate that the higher-order terms are negligible (which typically is the case in the delocalized regime, as we discuss below). To distinguish these different orders it will be useful to introduce the quantities $O_1=1-\gamma-\overline{\xi}_2/4$, $O_2=O_1+(\overline{\xi}_3/9+\overline{\xi}_{22}/16)$ and $O_3=O_2-(11\overline{\xi}_4/96+ \overline{\xi}_{32}/6+ \overline{\xi}_{222}/16)$. 

\begin{figure}[t]
\includegraphics[width=0.95\linewidth]{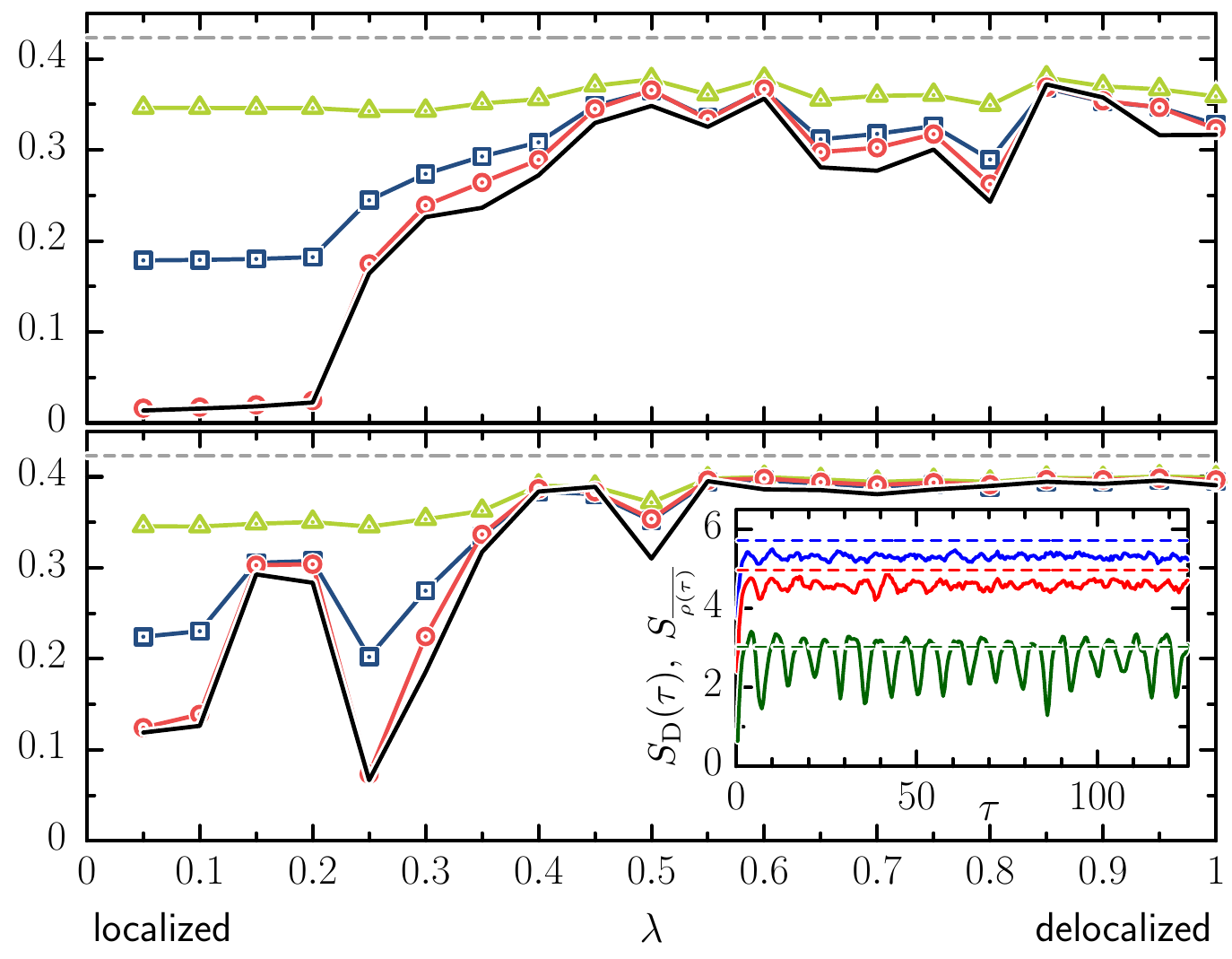} 
\caption{$\Delta S$ (solid black line) as a function of the coupling strength $\lambda$ for the DM with $j=20$, $N_t=250$, $\tau=10^7$, $\Delta\tau=250$, and quench strength $\delta \lambda=0.1$. The initial state is $\ket{n_0}=\ket{10}$ (top) and $\ket{100}$ (bottom). Dashed horizontal line corresponds to $1-\gamma$; symbols are orders $O_1$ (squares), $O_2$ (triangles)  and $O_3$ (circles). Inset: examples of  $\SD(\tau)$ as a function of $\tau$, with $\lambda=0.65$, $\delta\lambda=0.1$,  for three different initial states: (bottom) $\ket{10}$; (middle) $\ket{100}$; (top) $\ket{250}$. The dashed horizontal lines mark the corresponding values of $\Sde$.
\label{fig:Dickeorder}}
\end{figure}
To test the consistency of our analytical results we study two different models. 
Both of them undergo a localization-delocalization transition when varying one parameter. 
The first model is the Dicke model (DM) \cite{Dicke54} describing the dipole interaction of a single mode of a bosonic field, of frequency $\omega$,  with $n_s$ two-level particles, with level splitting $\omega_0$. The corresponding Hamiltonian is
\begin{equation}
H(\lambda)=\omega_0 J_z +\omega a^\dagger a+\frac{\lambda}{\sqrt{2 j}}(a^\dagger + a)(J_{-}+J_{+}).
\end{equation}
The collective angular momentum operators $J_z$, $J_\pm$ correspond to a pseudospin $j=n_s/2$,  and $a^\dagger$ ($a$) are 
creation (annihilation) operators of the field.
The DM undergoes a superradiant quantum phase transition in the thermodynamic limit ($n_s\to \infty$) at $\lambda_c=\sqrt{\omega_0\omega}/2$ \cite{dicke}. For finite $n_s$, there is a transition from Poissonian to Wigner-Dyson level spacing statistics at $\lambda\approx \lambda_c$. This marks a transition from quasi-integrability at small $\lambda$ to quantum chaos at large $\lambda$, as is verified using a semiclassical model in \cite{EmaryBrandes2003}. In our calculations we consider $\omega=\omega_0=1$ ($\lambda_c=0.5$) and the quench is implemented by changing $\lambda\to \lambda+\delta\lambda$. 

\begin{figure}[t]
\includegraphics[width=0.95\linewidth]{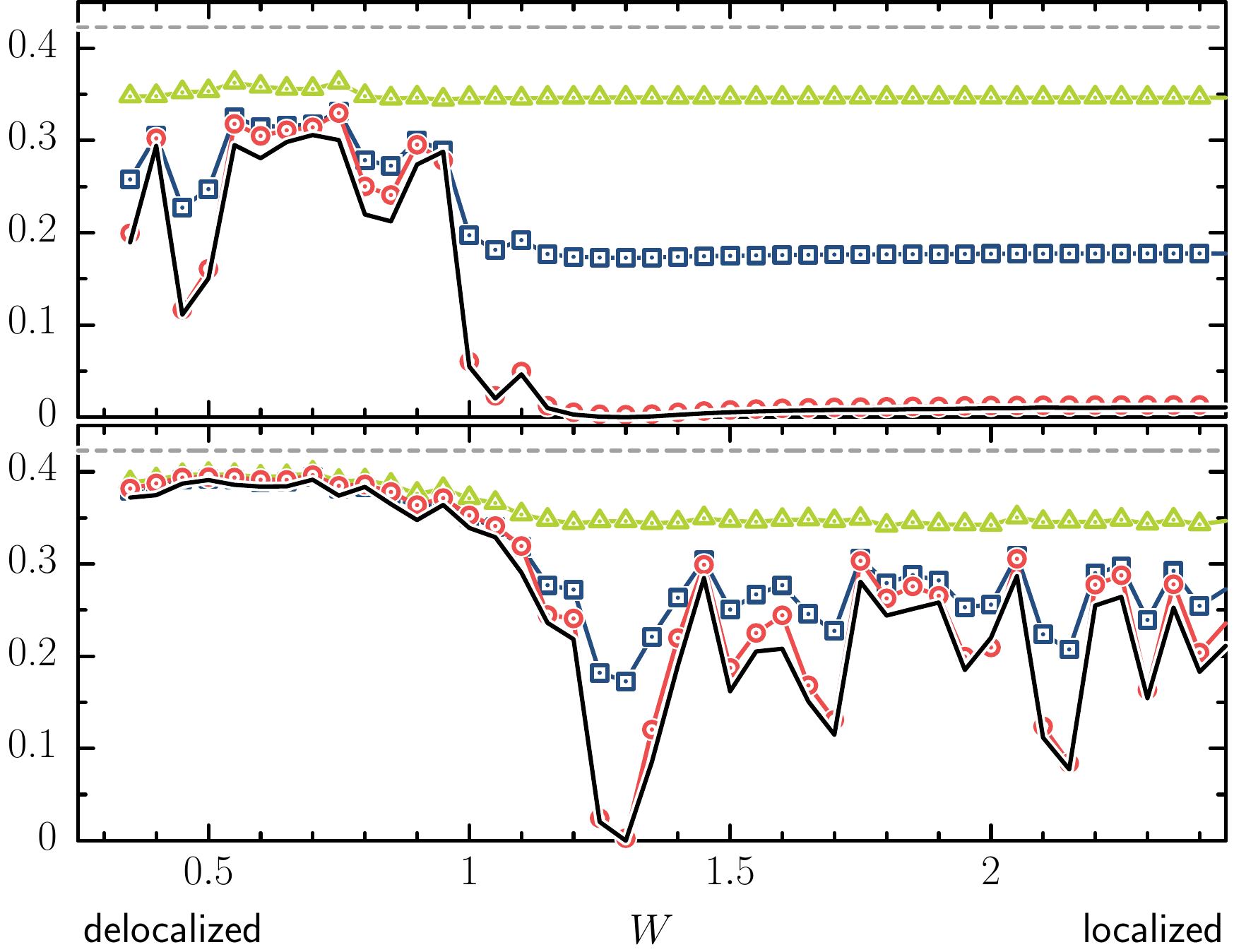} 
\caption{$\Delta S$ (solid black line) for the SW model with $p=0.06$, $\delta W=0.3$, $N=2^9$, $\tau=10^6$, $\Delta \tau=2500$ (top) and $\Delta \tau=3500$ (bottom), for one realization of disorder and of the shortcut links. Initial states are $\ket{n_0}=\ket{10}$ (top) and $\ket{200}$ bottom. Dashed horizontal line corresponds to $1-\gamma$; symbols are orders $O_1$ (squares), $O_2$ (triangles)  and $O_3$ (circles). 
\label{fig:SWorder}}
\end{figure}
The second model is a quantum smallworld (SW) system with disorder \cite{PhysRevB.62.14780, Giraud2005}. This is a one-dimensional tight-binding Anderson model having $N=2^{n_r}$ sites, with nearest-neighbor interaction and periodic boundary conditions, to which $p\,N$ shortcut links between sites are added, connecting $p\,N$ random pairs of vertices. The Hamiltonian which describes the quantum version of this system is  
\begin{equation}
H=\sum_{i}\varepsilon_i\op{i}{i}+\sum_{\langle i,j\rangle}V \op{i}{j}
+\sum_{k=1}^{\lfloor pN \rfloor} V(\op{i_k}{j_k}+\op{j_k}{i_k}),
\end{equation}
where $\varepsilon_i$ are Gaussian random variables with zero mean and width $W$, and $V=1$. The second sum runs over nearest-neighbors, while the third term describes the shortcut links of smallworld type, connecting random pairs $(i_k,j_k)$. When $p=0$ the model coincides with the usual one-dimensional Anderson model where all states are known to be localized with localization length $l\sim1/W^2$ for small $W$ \cite{Kramer1993}. The presence of long-range interacting pairs for $p>0$ induces a delocalization transition from localized states for large $W$ to delocalized states at small $W$ \cite{Giraud2005, NOUSunpub}.
The quench in this case is implemented by keeping the shortcut links $(i_k,j_k)$ fixed and changing $W\to W-\delta W$ . 

To compute $\Delta S$ we fully diagonalize $H$ and $H'$ to calculate the $h_n$ and perform the average \eqref{DeltaS} over a window $[\tau,\tau+\Delta \tau]$ for a very large $\tau$. Examples of $\Sde$ and $\overline{\SD(\tau)}$ are plotted in the inset of Fig.~\ref{fig:Dickeorder}. In the case of the DM, we take into account the parity symmetry, and we truncate the phonon basis to a finite size $N_t$  (we only consider energies well inside the converged part of the spectrum).

In Fig.~\ref{fig:Dickeorder} we show $\Delta S$ for the DM as a function of the coupling strength $\lambda$ for two different initial states. 
At large $\lambda$ (delocalized regime), order $O_1$ already approximates $\Delta S$ rather well. At small $\lambda$ (localized regime), $O_1$ and $O_2$ tend to a constant value corresponding to the fully localized case $\overline{\xi}_\mu\to 1$, while $O_3$ matches $\Delta S$ very well. The higher the energy, the smaller the localized plateau is, as illustrated in Fig.~\ref{fig:Dickeorder}; this can be understood by the fact that the quench induces more transitions at higher energies (see also \cite{GarciaMata2015}). In Fig.~\ref{fig:SWorder} we present similar results for the SW model: in the delocalized (small-$W$) regime, order $O_1$ gives again a good approximation of $\Delta S$, while in the localized regime $O_3$ gives a very good approximation for $\Delta S$. Note that Fig.~\ref{fig:SWorder} corresponds to a single disorder and shortcut link realization (we have checked that the results are equivalent for any realization with sufficiently large $\Delta \tau$). 

In order to understand these features, we  consider two limiting situations. When equilibrium is reached `ideally', the eigenvectors $\ket{m}$ of $H'$ are uniformly spread (i.~e.~delocalized) in the $\ket{n}$ basis, thus each overlap is $|\langle m\ket{n}|^2\sim 1/N$, so that $c_{mn}\sim 1/N^2$ for all $m,n$. For the PR this implies $\xi_{q,n}\sim N^{1-q}$, and thus $\overline{\xi}_{\mu}\sim N^{\textrm{lg}(\mu)-|\mu|}$, with lg$(\mu)$ the number of terms in the sequence $\mu$. In the opposite case where eigenvectors $\ket{m}$ and  $\ket{n}$ coincide, all $\overline{\xi}_{\mu}$ are equal to 1. These distinct features reflect in the behavior of the moments $\overline{h_n^k}$ (for instance, the sum of mean return probabilities $\sum_{n_0}\overline{h_{n_0}}$ was proposed as a tool to measure the degree of equilibration of an isolated quantum system \cite{Luck2015}). 

In these two extreme regimes $\Delta S$ has very distinct behaviors. In the delocalized case where $\overline{\xi}_{\mu}\sim N^{\textrm{lg}(\mu)-|\mu|}$, we have for instance $\overline{\xi}_2\sim 1/N$, while the two last brackets in Eq.~\eqref{conjecture3} correspond to terms scaling as $1/N^2$ and $1/N^3$, respectively. One can thus truncate the sum \eqref{bigsum} to any order by keeping terms with the same power in $N$. Order 0 is simply given by the constant $1-\gamma$, which coincides with the result of \cite{Ikeda2015}. Keeping the term in $1/N$ yields our first main result Eq.~\eqref{conjecturenext}, and as already mentioned explains the conjecture of \cite{Ikeda2015}.

In the localized case, on the other hand, the expansion \eqref{bigsum} is no longer valid, as each term $\overline{\xi}_{\mu}$ is of order 1. However, the truncation \eqref{conjecture3} happens to yield a very good approximation for $\Delta S$, as shown on the physical models in Figs.~\ref{fig:Dickeorder} and \ref{fig:SWorder}, where at small $\lambda$ or large $W$ the numerically computed $\Delta S$ is almost indistinguishable from the expression for $O_3$. This can be understood via the following perturbation-theory approach. Let us consider the simplest case of a localized model, where $H'=H+\epsilon V$ and $V$ is a symmetric matrix with elements of order 1. Standard perturbation theory for $\epsilon\ll 1$ yields 
\begin{equation}
\label{overlap_perturb}
\langle n\ket{m}=\delta_{nm}\left(1-\frac{1}{2}\sum_{k\neq m}v_{km}^2\right)+\epsilon(1-\delta_{nm}) v_{nm},
\end{equation}
with $v_{nm}=\epsilon V_{nm}/(E_m-E_n)$ and $\delta_{nm}$ is the Kronecker symbol. Inserting this expression into \eqref{Unn0}, we get, at lowest order in $\epsilon$,
\begin{equation}
\label{hnperturb}
h_n=\left\{
\begin{array}{cc}4\sin^2[(E'_n-E'_{n_0})\tau/2]v_{nn_0}^2\quad&n\neq n_0\\
&\\
1-2\sum_{k\neq n_0}v_{n_0k}^2\quad&n=n_0\,.
\end{array}
\right.
\end{equation}
Upon averaging over $\tau$ in \eqref{DeltaS}, the term $n=n_0$ does not contribute (as it does not depend on $\tau$), while terms $n\neq n_0$ yield a contribution involving the average $\overline{z\log z}-\overline{z}\log(\overline{z})=\frac12(1-\log 2)$ (where $z\stackrel{\rm def}{=}\sin^2x$).
At order $\epsilon^2$ the average \eqref{DeltaS} is thus given by
\begin{equation}
\label{deltaSperturb1}
\Delta S\simeq2(1-\log 2)\sum_{n\neq n_0}v_{nn_0}^2.
\end{equation}
One can similarly calculate a perturbation expansion for the $\overline{\xi}_{\mu}$, by injecting \eqref{overlap_perturb} into $c_{mn}$ given by \eqref{hnbar} and calculating the PR defined by \eqref{xiqn}. At order $\epsilon^2$ one gets 
\begin{eqnarray}
\label{xiperturb}
\overline{\xi}_2&\simeq&1-\sum_{n\neq n_0}v_{nn_0}^2,\quad
\overline{\xi}_3\simeq\overline{\xi}_{22}\simeq1-\frac32\sum_{n\neq n_0}v_{nn_0}^2,\nonumber\\
\overline{\xi}_4&\simeq&\overline{\xi}_{32}\simeq\overline{\xi}_{222}\simeq1-\frac74\sum_{n\neq n_0}v_{nn_0}^2.
\end{eqnarray}
Using \eqref{xiperturb} to calculate the successive orders of $\Delta S$ given in \eqref{conjecture3} we obtain
\begin{eqnarray}
\label{deltaSperturb2}
O_1\simeq&\frac34-\gamma+\frac14\sum_{n\neq n_0}v_{nn_0}^2,\\
\label{deltaSperturb3}
O_2\simeq&\frac{133}{144}-\gamma-\frac{1}{96}\sum_{n\neq n_0}v_{nn_0}^2,\\
\label{deltaSperturb4}
O_3\simeq&\frac{167}{288}-\gamma+\frac{227}{384}\sum_{n\neq n_0}v_{nn_0}^2.
\end{eqnarray}
As the constant in front of $\epsilon^2$ in \eqref{deltaSperturb3} is $1/96$, at this order $\Delta S$ is essentially equal to $133/144-\gamma\simeq 0.346395$, as can be seen in the figures in the localized region. At the next order $O_3$ on the other hand, the constant $167/288-\gamma\simeq 0.00264545$ almost vanishes, while $227/384\simeq 0.591146$ is numerically very close to $2(1-\log 2)\simeq 0.613706$. Thus the expression \eqref{deltaSperturb4} almost coincides with \eqref{deltaSperturb1}, which explains why the truncation at $O_3$ works so well. Truncation at order 4 (which can be obtained from the general result given explicitly in the Supplemental material) would yield $\Delta S\simeq 1.2184 - 1.68839 \sum_{n\neq n_0}v_{nn_0}^2$, so that this approximation is worse than order 3 (and the same happens at higher orders). $O_3$ thus appears as the optimal truncation in the localized regime. As was noted above, this truncation also works quite well in the delocalized regime. 
One sees from the numerical results that Eq.~\eqref{conjecture3} is in fact a good approximation for $\Delta S$ over the whole range of parameters.

To summarize, 
           there is a quest for a quantum entropy with all the required properties.
The DE is a good candidate and here we give an analytical expression for its time average as a function of eigenvector properties only, to a very good approximation and independently of the dynamical properties, in particular of the localization-delocalization or chaotic-integrable characteristics.
Our results establish the validity of the conjecture of Ref.~\cite{Ikeda2015} and extend its accuracy. 
These results are
relevant in the relation between many-body localization transition and the study of equilibration in non-equilibrium isolated quantum systems.

\acknowledgments
IGM and OG received partial funding from a binational project from  CONICET (grant no. 1158/14) and CNRS (grant no PICS06303). 
IGM also received a funding from Universit\'e Toulouse III Paul Sabatier as an invited professor.
IGM thanks D. A. Wisniacki for fruitful discussions.

%

\onecolumngrid
\newpage
\appendix
\begin{center}
{\large\textbf{Supplemental material for  ``Average diagonal entropy in non-equilibrium isolated quantum systems
''}}
\end{center}

In this Supplemental material we provide a detailed proof of the general relation Eq.~(14) of the main text. We recall that $\Delta S$ is defined as
\begin{equation}
\label{DeltaS2}
\Delta S=-\sum_n \overline{h_n}\ln\overline{h_n}+\sum_n \overline{h_n\ln h_n},
\end{equation}
with transition probabilities given by
\begin{equation}
\label{hnbis2}
h_n=\!\!\sum_{m_1,m_2}\!e^{i(E'_{m_1}-E'_{m_2}) \tau}\langle n\ketbra{m_2}n_0\rangle\langle n_0\ketbra{m_1}n\rangle.
\end{equation}
To make notations lighter we will drop the index $n$ until the very last equation. The demonstration goes as follows. First we obtain an expression for the moments $\overline{h^k}$ as sums over integer partitions (section \ref{subs1}). 
Then, through the inverse Laplace transform, we can get the distribution $P(h)$ to calculate \equa{DeltaS} (section \ref{subs2}). 
To achieve this we first have to reexpress the moments in a way where resummation and simplification becomes feasible (section \ref{subs3}), and finally obtain the full expression for $\Delta S$ (section \ref{subs4}).

\subsection{Moments as a sum over integer partitions}
\label{subs1}
The first step is to calculate the moments $\overline{h^k}$ averaged over $\tau$, assuming that the quantities 
$e^{ \left[i\left(\sum_{i=1}^k E'_{m_{2i-1}}-\sum_{i=1}^{k} E'_{m_{2i}}\right)\tau\right]}$ average to 1 if the sets $\{m_{2i-1},1\leq i\leq k\}$ and $\{m_{2i},1\leq i\leq k\}$ are permutations of each other, and to 0 otherwise. In the main paper we gave the first averages
\begin{eqnarray}
\label{moment2}
\overline{h^2}/\overline{h}^2&=&2-\xi_2\\
\label{moment3}
\overline{h^3}/\overline{h}^3&=&6-9\xi_2+4\xi_3\\
\label{moment4}
\overline{h^4}/\overline{h}^4&=&24-72\xi_2+18(\xi_2)^2+64\xi_3-33\xi_4
\end{eqnarray}
in terms of the participation ratios (PR) of the vectors $(c_{m})_{m}$ for integer $q$,
\begin{equation}
\label{xiqn2}
\xi_{q}\equiv\frac{\sum_m c_{m}^q}{(\sum_m c_{m})^q}, \qquad c_{m}=|\langle m|n_0\rangle|^2|\langle m|n\rangle|^2, \qquad \sum_mc_m=\overline{h}.
\end{equation}
It is possible to obtain these expressions in a systematic way by introducing integer partitions. It is usual to denote by $\lambda\vdash k$ a partition $\lambda=(\lambda_1,\lambda_2,\ldots)$ of $k$, with $\lambda_1\geq\lambda_2\geq\ldots$ and $\sum_i\lambda_i=k$ (it can be padded on the right by an arbitrary number of zeros). The products of $\xi_q$ appearing in the expressions \eqref{moment2}--\eqref{moment4} for $\overline{h^k}$ correspond to all possible integer partitions of the integer $k$. For instance, noticing that $\xi_1=1$, the terms $\xi_4$, $\xi_3\xi_1$, $\xi_2^2$, $\xi_2\xi_1^2$ and $\xi_1^4$ contributing to Eq.~\eqref{moment4} correspond to the partitions $4=3+1=2+2=2+1+1=1+1+1+1$. Following standard textbooks \cite{Macdo}, one can define several families of symmetric polynomials of the variables $c_m$, $1\leq m\leq N$. We set
\begin{equation}
\label{mlambda}
m_{\lambda}=\sum_\sigma c_1^{\sigma(\lambda_1)}c_2^{\sigma(\lambda_2)}\ldots c_N^{\sigma(\lambda_N)},
\end{equation}
where the sum runs over all permutations of $(\lambda_1,\lambda_2,\ldots,\lambda_N)$ (see p.~18 of \cite{Macdo}); if $N$ is smaller than the number of nonzero $\lambda_i$'s then by convention $m_{\lambda}=0$. We also define the polynomials
\begin{equation}
\label{plambda}
p_{\lambda}=(\sum_m c_m^{\lambda_1})(\sum_m c_m^{\lambda_2})\ldots
\end{equation}
(see p.~24 of \cite{Macdo}). The $p_{\lambda}$ are simply related to the PR defined by Eq.~\eqref{xiqn2} by
\begin{equation}
\label{pxi}
p_{\lambda}=\overline{h}^k\xi_{\lambda_1}\xi_{\lambda_2}\cdots
\end{equation}
For $\lambda\vdash k$, the $p_{\lambda}$ and $m_{\lambda}$ are related by the linear relation
\begin{equation}
\label{pLm}
p_{\lambda}=\sum_{\mu\vdash k}L_{\lambda\mu}m_{\mu}
\end{equation}
(p.~103 of \cite{Macdo}), with $L_{\lambda\mu}$ an invertible lower-triangular matrix of integers indexed by partitions $\lambda,\mu$, of $k$.

Taking the $k\,$th power of Eq.~\eqref{hnbis2} and keeping only terms where the energies exactly compensate, we get the general expression of the $k\,$th moment as
\begin{equation}
\label{hk1}
\overline{h^k}=\sum_{m_1,\ldots,m_k}P(m_1,\ldots,m_k)c_{m_1}c_{m_2}\ldots c_{m_k}
\end{equation}
with $P(m_1,\ldots,m_k)$ the number of permutations of the $m_i$. We then group together all terms with the same `pattern' of indices, e.g., for $k=3$, terms with $m_1=m_2=m_3$ or $m_1=m_2\neq m_3$ or $m_1\neq m_2\neq m_3$, which corresponds to the different integer partitions 3=2+1=1+1+1. The expression \eqref{hk1} simply reduces to
\begin{equation}
\label{hk2}
\overline{h^k}=\sum_{\lambda\vdash k}P_{\lambda}^2m_{\lambda}
\end{equation}
with $P_{\lambda}$ the multinomial coefficient associated with $(\lambda_1,\lambda_2,\ldots)$ and $m_{\lambda}$ the symmetric polynomial \eqref{mlambda}. In order to relate the mean moments to the PR, we want to express $\overline{h^k}$ by means of the $p_{\lambda}$ rather than the $m_{\lambda}$; inverting the relation \eqref{pLm} we get
\begin{equation}
\label{hk3}
\overline{h^k}=\sum_{\lambda\vdash k}\sum_{\mu\vdash k}P_{\lambda}^2(L^{-1})_{\lambda\mu}p_{\mu}.
\end{equation}
Setting $Z_{\mu}=\sum_{\lambda\vdash k}P_{\lambda}^2(L^{-1})_{\lambda\mu}$ we have the final compact expression
\begin{equation}
\overline{h^k}=\sum_{\mu\vdash k}Z_{\mu}p_{\mu}.
\label{finalhk}
\end{equation}
The $L_{\lambda\mu}$, and thus the $Z_{\mu}$, can be calculated very easily using mathematical software, while $p_{\mu}$ is obtained from Eq.~\eqref{pxi}. For instance, one gets for $k=3=2+1=1+1+1$ (where partitions are ordered in that reverse lexicographic order) the matrix $L=\{\{1,0,0\},\{1,1,0\},\{1,3,6\}\}$, and multinomial coefficients $P_{3}=1$, $P_{21}=3$ and $P_{111}=6$, so that $Z_{\mu}$ is the vector $(4, -9, 6)$, which allows to recover indeed Eq.~\eqref{moment3}.

\subsection{From moments to entropy difference}
\label{subs2}
The knowledge of the moments \eqref{finalhk} allows to reconstruct the probability distribution $P(h)$. Indeed, let
\begin{equation}
\label{defM}
M(t)=\int_{0}^{\infty}dh P(h) e^{t h}
\end{equation}
be the moment generating function of $P(h)$. Then $M(t)$ can be expressed as a series
\begin{equation}
\label{sumM2}
M(t)=\sum_{k=0}^{\infty}\frac{\overline{h^k}}{k!}t^k.
\end{equation}
It is then possible to get $P(h)$ from inverse Laplace transform of $M(t)$, namely
\begin{equation}
\label{inverselaplace}
P(h)=\frac{1}{2i\pi}\int_{c-\textrm{i}\infty}^{c+\textrm{i}\infty}dt e^{t h}M(-t)
\end{equation}
with $c$ a real number such that the contour in \eqref{inverselaplace} goes to the right of all poles (the $M(-t)$ comes from the fact that \eqref{defM} is not exactly a Laplace transform as there is a factor $e^{t h}$ rather than $e^{-t h}$). The entropy difference $\Delta S$ associated with $h$ is then simply given by
\begin{equation}
\label{entropydiff}
-\overline{h}\ln\overline{h}+\overline{h\ln h}=-\overline{h}\ln\overline{h}+\int_{0}^{\infty}dh P(h) h\ln h.
\end{equation}
In order to perform the sum over $k$ in \eqref{sumM2} we first have to rewrite the moments $\overline{h^k}$ under a form where the dependence on $k$ is more transparent.

\subsection{Reexpressing the moments}
\label{subs3}
The aim of this section is to show that $Z_{\mu}$ appearing in Eq.~\eqref{finalhk} can be put under the form
\begin{equation}
\label{zmutilde}
Z_{\mu}=k!\,\binom{k}{s}\tilde{Z}_{\mu'}
\end{equation}
where $s=\sum_{\mu_i\geq 2}\mu_i$ and $\mu'$ is the partition of $s$ obtained by removing the 1's from $\mu$, with $\tilde{Z}_{\mu'}$ rational numbers. For instance, for all partitions of the form $\mu=(211\ldots1)$ we have $\mu'=(2)$ and $s=2$. From its explicit definition $Z_{\mu}=\sum_{\lambda\vdash k}P_{\lambda}^2(L^{-1})_{\lambda\mu}$ it is easy to calculate $Z_{2}=-1$, $Z_{21}=-6$, $Z_{211}=-12$, $Z_{2111}=-20$ and so on, so that Eq.~\eqref{zmutilde} holds with $\tilde{Z}_{2}=-\frac12$. Similarly we can compute $\tilde{Z}_{3}=\frac23$, $\tilde{Z}_{4}=-\frac{11}{8}$, $\tilde{Z}_{22}=\frac34$. In the special case of partitions of the form $\mu=(111\ldots1)$, Eq.~\eqref{zmutilde} can still be satisfied by setting $\mu'=(0)$ and $\tilde{Z}_{0}=1$. 

In order to show Eq.~\eqref{zmutilde}, we start by noting that the $L_{\lambda\mu}$ can be interpreted as the number of ways of adding together the $\lambda_i$ in order to obtain the $\mu_i$ (see \cite{Macdo} p.~103). For example, for $k=3$ the partition $\lambda=(111)$ can yield the partition $\mu=(21)$ by adding two 1's. There are three different ways of doing so, hence the entry $L_{111,21}=3$ given in subsection \ref{subs1}. 

Let $\lambda\vdash k$ be a fixed partition of $k$ such that all $\lambda_i\geq 2$. From their definition, the quantities $Z_{\mu}$ verify
\begin{equation}
\label{sum1}
\sum_{\mu\vdash k}L_{\mu\lambda}Z_{\mu}= P_{\lambda}^2.
\end{equation}
It can be rewritten as a double sum over the number $r$ of 1's in the partition $\mu$ and over partitions $\mu'$ of $k-r$ not containing any 1. We thus have
\begin{equation}
\label{sumrmup}
\sum_{r=0}^{k}\sum_{\genfrac{}{}{0pt}{}{\mu'\vdash k-r}{\mu'_i\geq 2}}L_{\mu\lambda}\tilde{Z}_{\mu}k!\,\binom{k}{r}= P_{\lambda}^2,
\end{equation}
where we have introduced the notation $\tilde{Z}_{\mu}=Z_{\mu}/(k!\,\binom{k}{r})$, and $\mu=(\mu'1\ldots 1)$ with $r$ numbers 1. The goal is to show that $\tilde{Z}_{\mu}$  in fact only depends on $\mu'$ and not on $r$.

If we adjoin a 1 to the partition $\lambda$, we get the partition $(\lambda 1)$, which is a partition of $k+1$. For this partition, Eq.~\eqref{sumrmup} yields
\begin{equation}
\label{sumkr}
\sum_{r=0}^{k+1}\sum_{\genfrac{}{}{0pt}{}{\nu'\vdash k+1-r}{\nu'_i\geq 2}}L_{\nu(\lambda 1)}\tilde{Z}_{\nu}(k+1)!\,\binom{k+1}{r}= P_{(\lambda 1)}^2=(k+1)^2P_{\lambda}^2.
\end{equation}
Since $L_{\nu(\lambda 1)}$ can be interpreted as the number of ways of adding together the $\nu_i$ in order to obtain the elements of $(\lambda 1)$, which are the $\lambda_i$ and the additional term 1, necessarily this additional 1 has to come from a 1 appearing among the $\nu_i$. In particular this implies that the term $r=0$ in the sum \eqref{sumkr} must vanish, and that $\nu$ is of the form $(\mu 1)$. There are $r$ possible ways of choosing this additional 1 (the total number of 1's in $\nu$), and then the number of ways to group the remaining $\nu_i=\mu_i$ to get the $\lambda_i$ is precisely $L_{\mu\lambda}$. Thus $L_{\nu(\lambda 1)}=L_{(\mu 1)(\lambda 1)}=r L_{\mu\lambda}$, and $\nu'=\mu'$. Shifting the sum in \eqref{sumkr} yields
\begin{equation}
\label{sumkr2}
\sum_{r=0}^{k}\sum_{\genfrac{}{}{0pt}{}{\mu'\vdash k-r}{\mu'_i\geq 2}}(r+1)L_{\mu\lambda}\tilde{Z}_{(\mu 1)}(k+1)!\,\binom{k+1}{r+1}=(k+1)^2P_{\lambda}^2,
\end{equation}
which after simplification reduces to
\begin{equation}
\label{sumkr3}
\sum_{r=0}^{k}\sum_{\genfrac{}{}{0pt}{}{\mu'\vdash k-r}{\mu'_i\geq 2}} L_{\mu\lambda}\tilde{Z}_{(\mu 1)}k!\,\binom{k}{r}=P_{\lambda}^2.
\end{equation}
Comparing Eq.~\eqref{sumkr3} with Eq.~\eqref{sumrmup} one gets that $\tilde{Z}_{(\mu 1)}=\tilde{Z}_{\mu}$. Proceeding in the same way recursively one can show that all $\tilde{Z}_{(\mu 1\ldots 1)}$ are equal: we denote them $\tilde{Z}_{\mu'}$, which proves Eq.~\eqref{zmutilde}.

\subsection{Resummation of contributions}
\label{subs4}
We are now in a position to calculate the distribution $P(h)$. Using the above result, Eq.~\eqref{finalhk} can be rewritten
\begin{equation}
\overline{h^k}=\sum_{\mu\vdash k}k!\,\binom{k}{s}\tilde{Z}_{\mu'}p_{\mu'}p_1^{k-s},
\label{finalhk2}
\end{equation}
using the definition \eqref{plambda} of $p_{\lambda}$ and the notation $\mu=(\mu'11\ldots 1)$, and with $s=\sum_i\mu'_i$. Any given sequence of numbers $\mu'=(\mu'_1\mu'_2\ldots)$ with $\mu'_i\geq 2$ will contribute to each moment $\overline{h^k}$ with $k\geq s$ through the partition $(\mu'11\ldots 1)$ of $k$ with $(k-s)$ 1's. The contribution of $\mu'$ to $M(t)$ will be (using $p_1=\overline{h}$)
\begin{equation}
\sum_{k=s}^{\infty}t^k\binom{k}{s}\tilde{Z}_{\mu'}p_{\mu'}p_1^{k-s}=\frac{t^s\tilde{Z}_{\mu'}p_{\mu'}}{(1-\overline{h}t)^{s+1}}.\label{mtgeneral}
\end{equation}
The inverse Laplace transform \eqref{inverselaplace} then yields the contribution of $\mu'$ to the probability distribution $P(h)$. There is a single pole of order $s+1$ at $-1/\overline{h}$, whose residue is given by
\begin{equation}
\label{bigsum2}
\frac{1}{s!}\frac{\tilde{Z}_{\mu'}p_{\mu'}}{\overline{h}^{s+1}}\lim_{t\to -1/\overline{h}}\frac{\partial^s}{\partial t^s}\left(t^se^{t h}\right)=
\sum_{r=0}^{s}\binom{s}{r}^2\frac{r!}{s!}\left(-\frac{h}{\overline{h}}\right)^{s-r}
\frac{e^{-h/\overline{h}}}{\overline{h}^{s+1}}\tilde{Z}_{\mu'}p_{\mu'}.
\end{equation}
The contribution to the entropy difference \eqref{entropydiff} is then obtained by evaluating integrals of the form
\begin{equation}
\int_{0}^{\infty}\!dh\left(-\frac{h}{\overline{h}}\right)^{a}\frac{e^{-h/\overline{h}}}{\overline{h}}h\ln h =(-1)^a (a+1)!\,\, \overline{h} \left(\ln\overline{h}+\frac{|S_{a+2}|}{(a+1)!}-\gamma\right),
\end{equation}
where $S_{a}$ are Stirling numbers of the first kind and $\gamma=0.5772\ldots\,$ is Euler's constant. Performing the summation over $r$ in \eqref{bigsum2}, this term reduces to $\overline{h}\left(\ln\overline{h}+1-\gamma\right)$ if $s=0$, and to
\begin{equation}
\label{bigsum3}
\frac{\tilde{Z}_{\mu'}p_{\mu'}}{\overline{h}^{s-1}}\sum_{r=0}^{s}\binom{s}{r}(-1)^r(r+1)\left(\ln\overline{h}-\gamma+\frac{|S_{r+2}|}{(r+1)!}\right)=\frac{\tilde{Z}_{\mu'}p_{\mu'}}{s(s-1)\overline{h}^{s-1}}
\end{equation}
if $s\geq 2$. The last expression has been obtained by using the identities for $s\geq 2$ (see e.g.~\cite{spivey})
\begin{equation}
\sum_{r=0}^{s}\binom{s}{r}(-1)^r(r+1)=0
\end{equation}
and
\begin{equation}
\sum_{r=0}^{s}\binom{s}{r}(-1)^r\frac{|S_{r+2}|}{r!}=\frac{1}{s(s-1)}.
\end{equation}
Summing up contributions from all possible $\mu'$ one finally has
\begin{equation}
\label{bigsum4}
-\overline{h}\ln\overline{h}+\overline{h\ln h}=\overline{h}\left(1-\gamma\right)+\sum_{\mu'}\frac{\tilde{Z}_{\mu'}\overline{h}}{s(s-1)}\frac{p_{\mu'}}{\overline{h}^{s}}
\end{equation}
where $s=\sum_{i}\mu'_i$ and the sum runs over all sequences $\mu'=(\mu'_1\mu'_2\ldots)$ with $\mu'_i\geq 2$. By increasing order these partitions are $(2),(3),(4),(22),(5),(32),\ldots$. Using Eq.~\eqref{pxi} the term $p_{\mu'}/\overline{h}^{s}=\xi_{\mu'_1}\xi_{\mu'_2}\ldots$ is just a product of generalized participation ratios. Reintroducing the dependence in $n$ for $h\equiv h_n$, we get (recall that $\sum_n\overline{h_n}=1$) the final expression
\begin{equation}
\label{finalexpr}
-\sum_n \overline{h_n}\ln\overline{h_n}+\sum_n \overline{h_n\ln h_n}=1-\gamma+\sum_{\mu'}\frac{\tilde{Z}_{\mu'}}{s(s-1)}\overline{\xi}_{\mu'}
\end{equation}
with $s=\sum_{i}\mu'_i$ and $\overline{\xi}_{\mu}\equiv\sum_n\overline{h_n}\,\xi_{\mu_1,n}\xi_{\mu_2,n}\ldots$.

\end{document}